\documentclass{PHYEAUTH}

\usepackage{amsmath}
\usepackage{color}
\usepackage{amssymb}
\usepackage{graphicx}
\usepackage{dcolumn}
\usepackage{bm}

\def\be{\begin{eqnarray}}
\def\ee{\end{eqnarray}}
\def\ba{\begin{array}}
\def\ea{\end{array}}

\begin{document}

\begin{frontmatter}

\title{Spatial Distribution of the Incompressible Strips at Aharonov-Bohm Interferometer}

\author[l1]{E. Cicek}
\author[l1]{, A. I. Mese}
\author[l2]{, M. Ulas}
\author[l3]{and A. Siddiki}

\address[l1]{Trakya University, Physics Department, Faculty of Arts and Science, 22030 Edirne, Turkey}
\address[l2]{Kirklareli University, Physics Department, Faculty of Arts and Science, Kavakli-Kirklareli, Turkey}
\address[l3]{Mu\~gla University, Physics Department, Faculty of Arts and Sciences, 48170-Kotekli, Mugla, Turkey}

\begin{abstract}
In this work, the edge physics of an Aharonov-Bohm interferometer
(ABI) defined on a two dimensional electron gas, subject to strong
perpendicular magnetic field $B$, is investigated. We solve the
three dimensional Poisson equation using numerical techniques
starting from the crystal growth parameters and surface image of
the sample. The potential profiles of etched and gate defined
geometries are compared and it is found that the etching yields a
steeper landscape. The spatial distribution of the incompressible
strips is investigated as a function of the gate voltage and
applied magnetic field, where the imposed current is confined to.
AB interference is investigated due to scattering processes
between two incompressible "edge-states".
\end{abstract}
\end{frontmatter}
%

The recent increasing interest towards the quantum Hall based
interferometers relies on the popularity of the quantum
information processing. In particular, the realization of electron
and quasi-particle interference experiments became a paradigm
~\cite{Goldman05:155313,goldman07:e/3}. A possible application is
proposed to use the non-Abelian 5/2 state for topological quantum
computation ~\cite{dassarma:quantum_comp}, which essentially has a
very similar structure of AB interferometers. The well established
experimental wisdom is that, to realize "clean" measurements are
extremely difficult, which strongly depends on the sample
geometry, crystal growth etc.
In typical AB interference experiments two propagating states are
brought to close vicinity, by the help of
gates~\cite{Goldman05:155313}. The edge states form a closed (or
almost closed) path, which enclosures certain amount of magnetic
flux. By changing the magnetic field or the area of the closed
path, one infers the phase of the particles. The conventional edge
picture is used to explain the observed AB oscillations
\cite{Chklovskii92:4026}, however, the actual distribution of the
edge-states is still not known for realistic samples, although,
several powerful techniques are used ~\cite{igor08:ab}. At the
recent experiments of Camino \emph{et. al}~\cite{goldman07:e/3},
show that the conventional AB theories, which neglect
electron-electron interactions, are unable to explain the
periodicity of the oscillations. The only very recent model of
Igor Zozoulenko~\cite{igor08:ab} could provide a reasonable
explanation for the unexpected behavior, which states that
"standard transport models" fail to understand the underlying
physics. However, the model geometry considered in their work is
quite different from the actual experimental setup. Here, we
provide an explicit calculation scheme to obtain the density and
potential profiles of an AB interferometer in the absence of an
external magnetic field and also under quantized Hall conditions.
Our calculation is based on solving the Poisson equation in 3D,
starting from the crystal growth parameters and the
lithographically defined surface patterns. A fourth order nearest
neighbor approximation is used on a square grid with open boundary
conditions and 3D fast Fourier transformation method is used to
obtain the solution iteratively
~\cite{Andreas03:potential,Sefa08:prb}. The outcome, \emph{i.e.}
the potential and electron density distributions, of this
calculation is used as an initial condition for the magnetic field
dependent calculations. The distinguishing part of our calculation
is that we do not have to assume only gate defined structures but
we can also handle etching defined geometries, which essentially
is the case for the experiments. We show that, the etching defined
samples present a sharper potential profile~\cite{Sefa08:prb},
therefore, the formation of the edge-states are strongly
influenced. This, obviously, effects the edge physics in
determining the AB oscillations.
\begin{figure}
{\centering
\includegraphics[width=.7\linewidth]{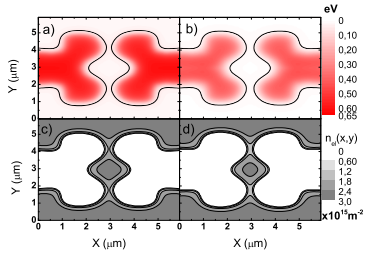}
%
\caption{ \label{fig:fig2}(a) and (c) considers only the gated
sample, whereas b) and d) is for both etched and trench gated
structure. a) and b) simulate the potential profiles of the sample
and the corresponding electron densities are shown in (c) and (d).
We ste the same gate voltage (-1.8 V, note that surface potential
is -0.7 V due to mid-gap pinning) for both situations. Potential
distribution of the gated sample presents smooth characteristic,
however, the potential profile of the trench gated structure is
very sharp. Hence the electron density at the island center is
higher in (d) compared to (c), in the island center. All the
figures are obtained at zero temperature and magnetic field.}}
\end{figure}

The simple description of an ABI, is such: let us assume a closed
path, a circle for simplicity, subject to a perpendicular magnetic
field with an area of $\pi r^2$ where $r$ is the radius. Since the
wave function travelling from one side should be the same as the
one the travelling from the other side, the phase difference of
the wave functions can be only integer multiples of the magnetic
flux quanta $\Phi_0=h/e$ encircled. In other words, for a given
flux $\Phi$ every spin resolved cyclotron orbit should satisfy the
condition $\Phi=BS_m=m\Phi_0$, where $m$ is the quantum number of
the orbit. Therefore, an orbit with a radius of magnetic length
$l=(\sqrt{\hbar/eB}~)$ will have an area $S_m=2\pi m l^2$,
resulting in $S_{m+1}-S_{m}=h/eB$. It is common to define the
occupation of these orbits by $\nu=n_{\rm el}h/eB$, where $n_{\rm
el}$ is the number of the electrons and this occupation is called
the filling factor. This picture should also hold for
non-interacting electrons which are confined by an external
potential $V_{\rm ext}(r)$, which essentially lifts the
degeneracy. In the experiments considered in this work, it was
shown that the number of electrons is fixed ($n_{\rm
el}=1700-2000$, varying with the sample size) in the "island",
which means that it is energetically almost impossible to add or
subtract an electron from the quantum dot region. If one keeps the
electron density fixed and decreases the $B$ field by a factor of
two, the filling factor increases by a factor of two, which
implies that the are enclosed by $\nu =1$ is now half.

In this work we only consider the ABI experiments conducted at the
integer quantize Hall regime, i.e. $\nu=n_{\rm el}/n_{B}=k$ and
$k=1,2,3,...$~\cite{Goldman05:155313} and the results reported in
Ref.\cite{goldman07:e/3} at the integer regime. The first work
concentrates the AB oscillations observed only at the $\nu=1$
plateau. In the above mentioned experiments, the magnetic field
$B$ and the front gate $V_{\rm FG}$ dependency of the AB period is
investigated and the results are discussed in terms of the Gelfand
and Halperin (GH)~\cite{Halperin94:etchedge} and Chklovskii,
Shklovskii and Glazman (CSG)~\cite{Chklovskii92:4026} edge models.
A hybrid formula is then used to describe the actual electron
density distribution \emph{non-self-consistently}. Procedure is as
follows: The GH model describes (almost) properly the etched edge
density profile and CSG model provides a $V_{\rm FG}$ dependency.
Therefore, the density distribution without gates is taken from
the GH model and its evolution depending on the front gate bias is
described by the CSG model. It is argued that, this hybrid model
is in agreement with the experiments with a difference of $13\%$
when comparing the surface area change $\Delta S_{\mu}$ of a
single \emph{edge-channel} at the gap $\mu$ as a function of
$V_{\rm FG}$. Such a relatively small difference, at a first
glance, looks impressive. However, in the second
report~\cite{goldman07:e/3}, where the results at $\nu=2$ is also
shown, it was stated that if the radius of the \emph{outer} ring
remains unchanged the AB oscillations can be explained.
\begin{figure}
{\centering
\includegraphics[width=0.6\linewidth]{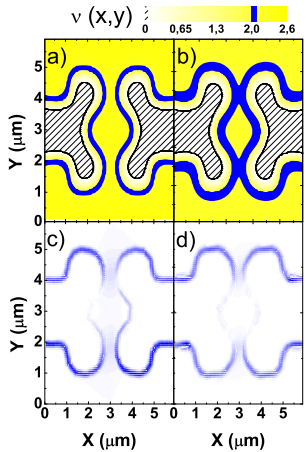}
\caption{ \label{fig:fig3} Spatial distribution of the
incompressible strips [(a) and (b)] and current distribution [(c)
and (d)] are shown at $B=$ 7.8 T [(a) and (c)] and 8.6 T [(b) and
(d)]. Blue highlight the ISs ($\nu =2$) and shaded areas indicate
the depleted regions in (a) and (b). Blue arrows indicate the
current distribution in (c) and (d). Fixed external current is
driven in $x$-direction where the 1D current density is set to be
$-1.06\times 10^{-4}$ A/m. Here we only consider the trench gated
structure (140 nm etched and $V_{\rm FG}=-1.8$ V).}}
\end{figure}

Our investigation is based on the calculation of the electron
density and screened potential self consistently within the
Thomas-Fermi approximation (TFA)~\cite{Sefa08:prb}. We consider
spinless electrons in the high magnetic field regime, consequently
the effective Hamiltonian reads,
\begin{equation}
H^{eff}=H_0+V_{\rm ext}(x,y)+V_{\rm int}(x,y).
\end{equation}
Here $H_0$ is the kinetic part, $V_{\rm ext}(x,y)$ and $V_{\rm
int}(x,y)$ are the external and interaction potentials,
respectively. The external potential is obtained by the above
mentioned 3D calculations considering the real experimental
structure, whereas the interaction potential (Hartree potential)
is calculated for a given density, boundary conditions and gate
pattern by solving the Poisson equation
\begin{equation}
V_{\rm int}(x,y)=\frac{2e^2}{\overline{\kappa}}\int
K(x,y,x',y')n_{\rm el}(x',y')dx'dy'
\end{equation}
where ${\overline{\kappa}}$ is the dielectric constant ($=12.4$
for GaAs). The Kernel $K(x,y,x',y')$ is the solution of the
Poisson equation preserving the periodic boundary conditions.
Spatial distribution of the electron density is obtained within
the TFA via,
\begin{equation}
n_{\rm el}(x,y)=\int dE D(E) f(E+V(x,y)-\mu^*),\label{eq:density}
\end{equation}
where $D(E)$ is the relevant (collision broadened) density of
states, $f(E,\mu^*, T)$ determines the particle statistics (Fermi
function) and $\mu^*$ is the electrochemical potential.  The total
potential $V(x,y)=V_{\rm int}(x,y)+V_{\rm ext}(x,y)$ and
eq.~\ref{eq:density} are calculated self-consistently to obtain
the electron and potential distributions at finite magnetic field
and temperature. We first compare the results of the gate and
trench gate defined samples. We show the potential profiles and
electron distributions both of these samples in
Fig.~\ref{fig:fig2}, the upper panels illustrate the potential
profiles and it is seen that the trench gate defined sample has a
sharper profile. Fig~\ref{fig:fig2}c and fig.~\ref{fig:fig2}d
represent the electron densities with gray scale, calculated for
two different definitions. These results are obtained at zero
magnetic field and temperature as an initial condition for our
non-zero magnetic field calculation. The maxima of the electron
density at the center of the q-dot matches perfectly ($<\%1$) with
the experimental value when considering the trench gated numerical
simulation.

Next we calculate the potential and electron density distribution
at finite temperature and $B$ field, only considering the trench
gated sample. In Fig.~\ref{fig:fig3} we show the spatial
distribution of the ISs at two different magnetic field values.
Shaded areas indicate that the electrons on the 2DEG are depleted
perfectly. The side-surface electrons which exists on the side
surface due to etching, yield larger shaded regions than the only
gated model. Since the potential is sharper at trench-gated
defined samples the widths of the incompressible strips are
narrower compared to gated samples. When considering the $B$ field
regime $2<\nu<4$ there is only a narrow interval where two
incompressible strips are close to each other
fig.~\ref{fig:fig3}a, however, do NOT merge ~\ref{fig:fig3}b. Fig.
~\ref{fig:fig3}c and fig.~\ref{fig:fig3}d presents the
corresponding calculated current distribution utilizing the local
Ohm's law~\cite{Guven03:115327,siddiki2004,Sefa08:prb}. We clearly
observe that, the imposed fixed current is confined within the
incompressible regions, where backscattering is absent. It is
known that the current flowing from the incompressible strips is
divergent free, thus it is not possible to inject current directly
to these strips. Hence, without scattering between two "edge
channels" one would not be able to observe any interference
pattern. The scattering mechanism comes from the impurities or the
electric field at the boundary between the ISs and compressible
region~\cite{SiddikiEPL:09}, which we implicitly include to our
calculations when calculating the local conductivity tensor
elements using the findings of self-consistent Born
approximation~\cite{siddiki2004}.

\begin{figure}
{\centering
\includegraphics[width=.5\linewidth]{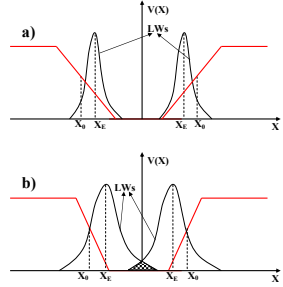}
%
\caption{ \label{fig:fig4} The (screened) potential and the shift
of the LWs for different magnetic fields. Slope of the potential
is small at higher magnetic field (a) compared to low $B$ value.}}
\end{figure}

We aim to investigate a situation where, the spin-degenerate groun
state Landau wave functions (LWs) overlap at a certain spatial
region where scattering matrix elements become finite, hence we
can observe the Aharonov-Bohm interference (ABI). Since the
screening is merely poor within the ISs, the total potential
exhibits a variation where the electron density is
constant~\cite{Siddiki03:125315}. Therefore an electric field
develops within this regions if we have a slope on the potential
profile, that can be computed from $
E_{x}=\frac{1}{e}\frac{\partial V(x)}{\partial x}$. Using this
equation and considering the calculated total potential profile on
can obtain the spatial shift, $X_e$, of the center coordinates
$X_0$ of the LWs from
\begin{equation}
X_e=X_0-\frac{eE_x}{(mw_c^2)},
\end{equation}
note that the LWs remain unaffected (i.e a Gaussian centered at
$X_0$) in the compressible regions, since the potential is
(almost) constant. Now, if the ISs are large and sufficiently far
apart, no scattering processes can take place hence no
interference can be observed. Such a case is shown in
fig.~\ref{fig:fig4}a. Whereas, if the two incompressible edge
states are close enough to each other and $E_x$ is large and LWs
overlap, \emph{i.e.} scattering takes place. This is shown in
Fig.~\ref{fig:fig4}b. If we apply a high magnetic field ISs become
broaden contrarily $E_{x}$ take small values so the shift of the
center coordinate is negligible and as a result no overlap occurs.
Since the widths of the incompressible strips are related to the
pattern geometry and crystal growth parameters, the observation of
AB interference patterns are extremely fragile. For such
experiments, one should design the sample geometry and choose the
magnetic field interval keeping in mind that the formation and the
spatial distribution of the incompressible strips are important.
The calculation of the actual scattering matrix elements for the
real structures considering self-consistently calculated wave
functions is beyond the scope of the present paper. However, we
have drawn the calculation scheme.

To summarize: We have calculated self-consistently the electron
and potential distribution considering the experimental sample
geometry and material properties. It is shown from the 3D
calculations that, the trench gated structure can be simulated in
a fairly good agreement (with $\%1$ error). In the second part of
our work we have also considered the interference conditions
depending on the magnetic field value and observed that the
scattering from one edge to the other is only possible if the two
incompressible strips are close enough, i.e. at the order of
magnetic length. If the two "edge states" merge no partition takes
place, hence interference pattern is smeared.

We would like to thank V. J. Goldman for fruitful discussions and
providing us the sample details. The Feza-G\"ursey Institute is
acknowledged for organizing the III. Nano-electronics symposium.
This work was supported by the Scientific and Technical Research
Council of Turkey (TUBITAK) for supporting under grant no 109T083.


\begin{thebibliography}{10}
\expandafter\ifx\csname url\endcsname\relax
  \def\url#1{\texttt{#1}}\fi
\expandafter\ifx\csname
urlprefix\endcsname\relax\def\urlprefix{URL }\fi
\expandafter\ifx\csname href\endcsname\relax
  \def\href#1#2{#2} \def\path#1{#1}\fi

\bibitem{Goldman05:155313}
F.~E. Camino, \emph{et.al}, Phys. Rev B (72) (2005) 155313.

\bibitem{goldman07:e/3}
F.~E. {Camino}, \emph{et.al}, Phys. Rev. Lett. 98~(7)
  (2007) 076805.

\bibitem{dassarma:quantum_comp}
S.~{Das Sarma}, \emph{et.al}, Phys. Rev. Lett. 94~(16) (2005)
166802.

\bibitem{Chklovskii92:4026}
D.~B. Chklovskii, \emph{et.al}, Phys. Rev. B (46) (1992) 4026.

\bibitem{igor08:ab}
S.~{Ihnatsenka}, \emph{et.al}, [cond-mat/mes-Hall 0803.4303].

\bibitem{Andreas03:potential}
A.~{Weichselbaum}, \emph{et.al}, Phys. Rev. E 68~(5) (2003)
056707.

\bibitem{Sefa08:prb}
S.~{Arslan}, \emph{et.al}, Phys. Rev. B 78~(12)
  (2008) 125423.

\bibitem{Halperin94:etchedge}
B.~Y. {Gelfand}, \emph{et.al}, Phys. Rev. B (49) (1994) 1862.

\bibitem{Guven03:115327}
K.~G{\"u}ven, \emph{et.al}, Phys. Rev. B (67) (2003) 115327.

\bibitem{siddiki2004}
A.~Siddiki, \emph{et.al} Phys. Rev. B (70) (2004) 195335.

\bibitem{SiddikiEPL:09}
A.~{Siddiki}, EPL (87) (2009) 17008--17014.

\bibitem{Siddiki03:125315}
A.~Siddiki, \emph{et.al}, Phys. Rev. B (68) (2003) 125315.

\end{thebibliography}
\bibliographystyle{elsarticle-num}

\end{document}